\documentclass[dvips,amsmath,amssymb]{article} 

\usepackage{icrctc07}

\title{Localising the H.E.S.S. Galactic Center point source} 

\shorttitle{H.E.S.S. Galactic Center location} 

\authors{C. van Eldik$^{1}$,
O.Bolz$^{1}$, I. Braun$^{1}$, G. Hermann$^{1}$, J. Hinton$^{2}$,
W. Hofmann$^{1}$.}  

\shortauthors{C. van Eldik et al.} 

\afiliations{$^1$Max-Planck-Institut f\"ur
Kernphysik, Saupfercheckweg 1, 69117 Heidelberg, Germany\\ $^2$School
of Physics \& Astronomy, University of Leeds, Leeds LS2 9JT, UK}
\email{Christopher.van.Eldik@mpi-hd.mpg.de}

\abstract{Observations by the H.E.S.S. system of
imaging atmospheric Cherenkov telescopes provide the most sensitive
measurements of the Galactic Centre region in the energy range 150 GeV
- 30 TeV. The vicinity of the kinetic centre of our galaxy harbours
numerous objects which could potentially accelerate particles to very
high energies (VHE, $> 100$~GeV) and thus produce the $\gamma$-ray flux 
observed. Within statistical and systematic errors, the centroid of
the point-like emission measured by H.E.S.S. was found 
\cite{Aharonian:2006wh} to be in good agreement with the position of the
supermassive black hole Sgr~A* and the recently discovered
PWN candidate G359.95-0.04 \cite{Wang:2005ya}. Given a 
systematic pointing error of about 30'', a possible association with
the SNR Sgr~A~East could not be ruled out with the 2004 \hess\ data.
In this contribution an update is given on the position of the
H.E.S.S. Galactic Centre source using 2005/2006 data. The systematic
pointing error is reduced to 6'' per axis using guiding telescopes for
pointing corrections, making it possible to exclude with high
significance Sgr A East as the source of the VHE \grs.}

\begin{document}

\newcommand{\gr}{$\gamma$-ray} \newcommand{\grs}{$\gamma$-rays}
\newcommand{\hess}{H.E.S.S.}  \newcommand{\hgc}{HESS~J1745-290}
\newcommand{\astar}{Sgr~A$^*$} \newcommand{\aeast}{Sgr~A~East}
\newcommand{\pwn}{G359.95-0.04}

\newcommand{\comment}[1]{\emph{(comment: #1)}}

\maketitle 

\section{Introduction} The centre of the Milky-Way is the most violent
and active region in our galaxy. Dust along the line of sight prevents
observations of the Galactic Centre (GC) by optical telescopes, but
precise data from this region have been obtained at radio, infrared,
X-ray, and hard X-ray/soft \gr\ ($<200$~keV) energies. These data have
established the 
existence of a $2.6\times 10^6\ M_{\odot}$ black hole at the kinematic
centre of our galaxy, commonly identified with the bright compact
radio source \astar, surrounded by a massive star cluster, a bright
supernova remnant shell, and giant molecular clouds (see, e.g., 
\cite{Melia07,Genzel:2007aa} for recent reviews).

VHE \gr\ emission from the direction of the Galactic Centre was
reported by several ground-based \gr\ observatories
\cite{Kosack:2004ri, Tsuchiya:2004wv, Aharonian:2004wa,
Albert:2005kh}. A recent deep exposure by \hess\
\cite{Aharonian:2006au} revealed the existence of two discrete VHE
\gr\ sources, on top of diffuse emission along the inner 300~pc of the
Galactic Centre ridge. One of the sources, HESS~J1747-281
\cite{Aharonian:2005br}, is identified with the pulsar wind nebula
(PWN) associated with the supernova remnant (SNR) G0.9+0.1. However,
no unique identification is possible for \hgc, the position of which
is within errors coincident with the kinematic centre of our galaxy.

A firm identification of \hgc\ is difficult because the GC region is
densely packed with sources of non-thermal radiation -- possibly
emitting at
VHE energies. In direct vicinity of the \hess\ source, at least three
different objects are discussed as possible counterparts of \hgc.
First, various models predict VHE \gr\ production near the
super-massive black hole itself (see, e.g., \cite{Aharonian:2005ti}).
\astar\ is partially 
surrounded by the bright, shell-like radio emission of the SNR \aeast\ 
\cite{Maeda:2005}, which is the second favoured candidate counterpart
of the VHE \gr\ emission. Finally, in a deep Chandra survey, \pwn, a
candidate pulsar wind nebula, was recently discovered
\cite{Wang:2005ya} only $8.7''$ away from \astar. Despite its faint
X-ray flux, models \cite{Hinton:2006zk} predict a TeV \gr\ flux that is
compatible with \hess\ observations. 

A precise localisation of \hgc\ is essential for shedding
light onto this source confusion. In this paper preliminary results
concerning a refined
position measurement of \hgc\ are reported using an improved telescope
pointing strategy, for which the systematic error on the observation
position is reduced by a factor of three compared to previous results.

\section{\hess\ observations of the Galactic Centre region} 

The most precise published results on the position of \hgc\ are
based on a 
50~h exposure carried out with the \hess\ array in 2004.  Within a
statistical error of 14'' the best-fit position of \hgc\ was found
\cite{Aharonian:2006wh} to coincide with the position of \astar. The
systematic pointing error of the \hess\ telescope system for this data
set is about 28'', already the most precise pointing in the field of
ground-based \gr\ astronomy.  

The results reported here are based on data recorded between May 14th
and July 27th, 2005, and between April 4th and September 24th, 2006.
The total good-quality exposure of the dataset is 73.2~h (live
time). Most of the 
data (66.1 h) were taken in ``wobble mode'' around \astar, i.e. the
observation direction was offset from the source direction by
typically $0.5^\circ-0.7^\circ$ in either right ascension or
declination. The remaining data were taken at various offsets, within
$1.4^\circ$ from \astar. The zenith angle distribution ranges from
$6^\circ-60^\circ$, and the mean zenith angle of observation is
$21.6^\circ$.

Data were analysed with the standard \hess\ calibration and
reconstruction chain \cite{crab}. \emph{Hard cuts}
\cite{Benbow:2005wj} were used for \gr\ selection, resulting in a
sample of well-reconstructed showers with an average angular
resolution of $0.07^\circ$ (68\% containment radius). 
The data show a strong excess of \grs\ from the direction of the GC
source \hgc, accompanied by diffuse \gr\ emission along the Galactic
Plane. An excess of 1300 $\gamma$~events is found within $0.1^\circ$
from the GC, corresponding to a statistical significance of 44.3 standard
deviations above background. The integral \gr\ flux above 1~TeV is in
agreement with published results based on 2004 data
\cite{Aharonian:2006wh}.

\section{Precision pointing} For an exact localisation of the centroid
of the VHE \gr\ emission, precise knowledge of the telescope pointing
direction is mandatory. The pointing deviation of individual telescopes is
typically of the order of 2-3'. Various causes have been identified,
with the most important ones being
small misalignments of azimuth and altitude axes during
construction,
sagging of telescope foundations over time,
(mostly) elastic deformations of the masts connecting the camera
body to the mirror dish,
gravitational bending of the mirror dish, and
inelastic deformations of the whole structure leading to
hysteresis effects.
The amount these effects contribute to the mispointing
strongly depends on the observation direction. It should however be
noted that - due to the rigidity of the steel construction - the overall
pointing deviation is very small given the size and weight of the \hess\
telescopes. 

\begin{figure}[htbp]
  \includegraphics[width=0.48\textwidth]{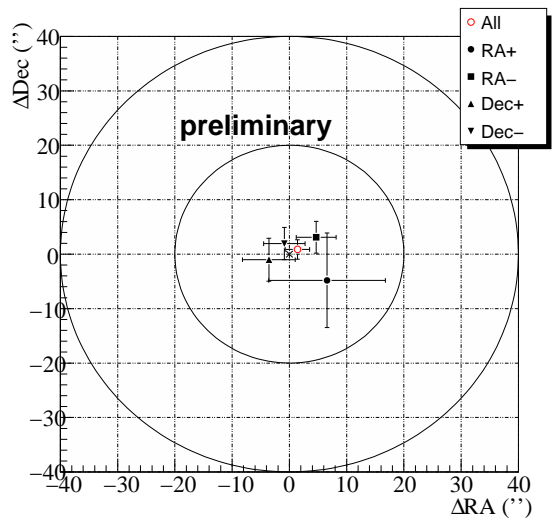}
  \caption{Position of the centroid of VHE \gr\ emission from
    PKS~2155-304 relative to its nominal position. Data were taken in 2006
    during an exceptional VHE \gr\ flare of this source
    \cite{benbow07}. The \gr\ excess was fit by a two-dimensional
    multi-gaussian profile representing the point spread function of the
    \hess\ instrument. The red data point shows the position derived from
    the full data set. When subdividing the data into the four wobble
    offsets, the positions shown by the black symbols are obtained. Note
    that for most of the RA+ wobble data, no bright stars were found in
    the field of view of the guiding telescopes, reducing the available
    live time for this analysis.}
  \label{fig:PointSourceTest}
\end{figure}

Most pointing deviations can be corrected for by taking calibration data at
regular intervals. Each telescope is pointed at typically 50 bright
stars uniformly distributed in the sky. The star is imaged by the
telescope mirror onto a screen in front of the Cherenkov camera, and
an image of the spot is recorded by a central CCD camera mounted at
the centre of the mirror dish. The position of each spot is then
compared to the nominal centre of the Cherenkov camera as determined
from eight positioning LEDs mounted on the camera body. The data are
fit with a 17 parameter model which accounts for elastic deformations
of the telescope structure. In the analysis of \gr\ data, this model
is then used to correct the position of the shower images in the focal
plane of the Cherenkov cameras. The precision achieved on the
observation direction of the \hess\ array is about 20'' per axis
\cite{Gillessen:2004tc}.

For the 2005-2006 data set presented here, the systematic error is
reduced 
further using guiding cameras mounted at each telescope. During
$\gamma$-ray observations, stars in the field of view
($0.3^\circ\times 0.5^\circ$) of these cameras are recorded at a
typical rate of 1~min$^{-1}$, and their reconstructed positions
matched to the Hipparcos and Tycho star catalogues. From this
information position-dependent corrections in right ascension and
declination are calculated for the individual \hess\
telescopes. Additionally, the position of the Cherenkov camera is
monitored by the central CCD camera. With this method, the systematic
error on the telescope orientation is reduced to 6'' per axis for
observations with the full \hess\ array (\cite{braun07}, details will
be published elsewhere).

The procedure was extensively tested on VHE $\gamma$-ray point sources
of known position. Fig. \ref{fig:PointSourceTest} shows a representative
study on the position of the high-frequency peaked BL Lac
PKS~2155-304. Excellent agreement with the nominal position of the
source is found even when splitting the data into different wobble
offsets.

\section{Position of \hgc} 
The position of \hgc\ is determined by
fitting, in a window of $\pm 0.2^\circ$ around the maximum excess, the
acceptance corrected and background subtracted \gr\ count map. 
Diffuse \gr\ emission is subtracted prior to the fit using the model
presented in \cite{Aharonian:2006au}.  The width of the 2-dimensional
gaussian fit to these data is composed of a fixed term
describing the mean angular resolution of the data set, and a
parameter left free to fit the intrinsic size of the source. The count
map is divided into sky bins of $0.04^\circ\times 0.04^\circ$, and the
fit function is integrated over the bin area for best
accuracy. $\chi^2$-minimisation is used to obtain the best-fit
position.

\begin{figure}[htbp]
  \includegraphics[width=0.52\textwidth]{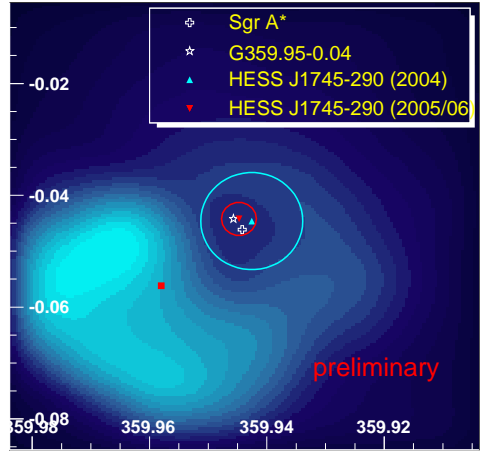}
  \caption{Smoothed 90~cm VLA radio image (reproduced from \cite{LaRosa00})
    of the SNR \aeast\ in Galactic coordinates. The
    position of \astar\ and \pwn\ are marked with a cross and a star,
    respectively. The blue triangle and circle mark the best fit position
    and total error (68\% CL) from the 2004 data set
    \cite{Aharonian:2006wh}. The best fit result of this analysis is shown
    by the red triangle and red circle. The red square marks the expected
    position of the centroid of the VHE \gr\ emission if it followed the
    observed radio flux of \aeast.}
  \label{fig:GCPosition}
\end{figure}

The best-fit position of \hgc\ in Galactic coordinates is $l=359^\circ
56'41.1''\pm 6.4''$~(stat.), $b=-0^\circ 2'39.2''\pm 5.9''$~(stat.).
These results are preliminary and subject to final checks.
Fig. \ref{fig:GCPosition} shows the new \hess\ position measurement on
top of a 90~cm VLA radio image of the inner 10~pc region of the
GC. The shell-like structure of the SNR \aeast\ is clearly visible.
The position of \hgc\ is coincident within $7.3'' \pm 8.7''$~(stat.)
$\pm8.5''$~(syst.) with the radio position of \astar\
\cite{saga_radio}, and is also consistent with the position reported from
the 2004 data set \cite{Aharonian:2006wh}. While the latter 
was marginally consistent with the radio emission from \aeast, the
result obtained in this analysis does rule out \aeast\ as the
counterpart of \hgc\ with high significance. Due to the improved
pointing accuracy of the \hess\ array, the probability that the
observed \gr\ flux is produced near the radio maximum of \aeast\ is
about $10^{-11}$. Assuming that the VHE \gr\ flux follows the radio
morphology of \aeast\ (corresponding to the red square in
Fig. \ref{fig:GCPosition}), the chance probability of finding the  
centroid of the emission at the reported position is $10^{-7}$. 

The position of \hgc\ agrees well with the location of the other two
counterpart candidates, \astar\ and \pwn, which are separated by only
$8.7''$. Since the pointing precision 
obtained in this work is at the limit of what can be achieved with
an instrument such as \hess, other measures have to be taken to
disentangle the remaining source confusion. The most promising method is
to search for variability in the VHE \gr\ flux, which would hint at
a connection between the VHE flux and \astar. The most convincing
signature would
be the detection of correlated flaring in X-rays and VHE \grs.
Such searches have been presented at this conference
\cite{vivier07, hinton07}.

\section{Acknowledgements} \small The support of the Namibian
authorities and of the University of Namibia in facilitating the
construction and operation of H.E.S.S. is gratefully acknowledged, as
is the support by the German Ministry for Education and Research
(BMBF), the Max Planck Society, the French Ministry for Research, the
CNRS-IN2P3 and the Astroparticle Interdisciplinary Programme of the
CNRS, the U.K. Science and Technology Facilities Council (STFC), the
IPNP of the Charles University, the Polish Ministry of Science and
Higher Education, the South African Department of Science and
Technology and National Research Foundation, and by the University of
Namibia. We appreciate the excellent work of the technical support
staff in Berlin, Durham, Hamburg, Heidelberg, Palaiseau, Paris,
Saclay, and in Namibia in the construction and operation of the
equipment.

\bibliography{proceeding_final}
\bibliographystyle{plain}

\end{document}